# Structure of Superconducting Ca-intercalated Bilayer Graphene/SiC studied using Total-Reflection High-Energy Positron Diffraction

Y. Endo*[1], Y. Fukaya[2], I. Mochizuki[3], A. Takayama[4], T. Hyodo[3] and S. Hasegawa[1]

[1] *Department of Physics, School of Science, The University of Tokyo,*

*7-3-1 Hongo, Bunkyo-ku, Tokyo 113-0033, Japan*

[2] *Advanced Science Research Center, Japan Atomic Energy Agency,*

*2-4 Shirakata, Tokai, Naka, Ibaraki 319-1195, Japan*

[3] *Institute of Materials Structure Science, High Energy Accelerator Research Organization (KEK),*

*Oho 1-1, Tsukuba, Ibaraki 305-0801, Japan*

[4] *Department of Physics, School of Advanced Science and Engineering, Waseda University,*

*3-4-1 Okubo, Shinjuku-ku, Tokyo 169-8555, Japan*

*Corresponding author. Tel: +81-3-5841-4209. E-mail: y.endo@surface.phys.s.u-tokyo.ac.jp

(Yukihiro Endo)

**Abstract**

We have investigated the atomic structure of superconducting Ca-intercalated bilayer graphene on a SiC(0001) substrate using total-reflection high-energy positron diffraction. By comparing the experimental rocking-curves with ones calculated for various structural models using a full-dynamical theory, we have found that Ca atoms are intercalated in the graphene-buffer interlayer, rather than between the two graphene layers. From transport measurements, the superconducting transition was observed to be at $T_c^{onset}$ = 4K for this structure. This study is the first to clearly identify the relation between the atomic arrangement and superconductivity in Ca-intercalated bilayer graphene.

**1. Introduction**

Graphene, a single atomic sheet of graphite, has attracted much attention because of the characteristic properties arising from its two-dimensional (2D) structure and the massless nature of the carriers therein [1, 2, 3]. Furthermore, superconductivity induced in graphene has been the subject of some focus [4, 5]. Much effort has been made to fabricate superconducting graphene by investigating the results of metal doping, following observation of the effect in bulk graphite intercalation compounds (GICs) [6, 7, 8, 9, 10]. Intensive research has been conducted to clarify the mechanism of superconductivity in bulk $C_6Ca$, which exhibits transition at $T_c$ = 11.5K, higher than any other GICs [11, 12]. Csányi *et al*. [11] claimed that electron doping from the intercalated Ca atoms induced superconductivity in graphite by making an otherwise unoccupied interlayer band occupied. It is known from X-ray diffraction [9] that the intercalated Ca atoms form a √3×√3 periodic structure and the carbon-layers align in AA stacking, and not AB stacking as in the pristine graphite.

Ca-intercalated bilayer graphene/SiC, which was regarded as the thinnest $C_6Ca$ structure, has also been studied. As with the bulk form, scanning tunneling microscopy and angle-resolved photoemission spectroscopy (ARPES) studies revealed that it also has √3×√3 periodicity [13]. Indication of superconducting transition was reported where it was observed that the resistance dropped steeply at 4 K, reaching zero at 2 K [5]. These studies were performed using bilayer graphene grown on a SiC(0001) substrate, which had a buffer layer beneath the graphene layers. There was, however, no direct evidence of the exact location of the intercalated-Ca atoms. To fully understand the superconducting mechanism of this compound, it is necessary to make a comprehensive structural analysis with atomic resolution.

In this study, we report on the results of investigations into the atomic structures of pristine bilayer graphene grown on SiC(0001), Ca-intercalated bilayer graphene showing superconducting transition at $T_c^{onset}$ = 4K, and Ca-desorbed bilayer graphene, using total-reflection high-energy positron diffraction (TRHEPD). By comparing rocking curves calculated for various structural models with the experimental data, we found that Ca atoms are intercalated in the graphene-buffer interlayer, rather than between the graphene layers.

## 2. Experimental procedure

The graphene was fabricated on an *n*-type Si-terminated 6*H*-SiC substrate by direct-current heating up to 1400°C under ultra-high vacuum (UHV) condition (~$10^{-9}$ torr). The graphene sample was transferred to a TRHEPD measurement chamber in KEK-IMSS [16] following angle-resolved photoemission spectroscopy (ARPES) measurements. We estimated the number of graphene sheets by ARPES [14, 15] and TRHEPD measurements. The band dispersion obtained by the ARPES measurements showed two parabolic $\pi$ bands, which are characteristic of bilayer graphene [as shown in figure S1 of Supplementary Data]. The results indicate that the dominant area of the sample is bilayer graphene. Furthermore, we checked the existence of some mono- and trilayer coverage by TRHEPD analysis; this is discussed later. All the processes for the subsequent fabrication of Ca-intercalated samples were performed in this UHV chamber.

As seen in Fig. 1(a), after heating up to 600°C to remove contaminants adsorbed on the surface, the reflection high-energy electron diffraction (RHEED) pattern clearly exhibited the fundamental spots (1×1) of the graphene layers and the SiC substrate and the superlattice spots (6√3×6√3) from the buffer layer between the graphene and SiC. The superlattice originates from the lattice mismatch between the SiC substrate and buffer layer with Si-C bonding [17, 18, 19]. Starting with Li atom intercalation, Ca deposition was accomplished following Li-Ca replacement as reported in a previous study [13]. Li atoms were deposited at room temperature using a Li dispenser (SAES Getters) under UHV (~$10^{-10}$ Torr). The resultant sample showed a characteristic √3×√3 RHEED pattern of Li-intercalated bilayer graphene [20]. Subsequently, it was annealed at ~180°C to desorb the excess Li atoms (forming 3D clusters) on top of the graphene. This heating process resulted in a much clearer RHEED pattern. Then, Ca-deposition was carried out using a custom-made Knudsen cell system; the cell was made of carbon. The Li-intercalated graphene sample was heated up to 200-270°C during the Ca-deposition. This temperature range was higher than the desorbing temperature of the excess Li atoms but lower than the disappearing temperature of √3×√3 spots from Li intercalation. Through the process, √3×√3 streaks and spots from some 3D clusters were observed simultaneously. Post annealing at 280-290°C made the 3D spots weaker or disappear.

This temperature range was slightly higher than the disappearing temperature of the √3×√3 spots from the Li-intercalation. By repeating the process of Li and Ca deposition-and-desorption a number of times, the √3×√3 streaks due to the Ca-intercalation became more defined [RHEED: Fig. 1 (b), TRHEPD: Fig. 1 (d)]. The RHEED pattern still showed the buffer layer's 6√3×6√3 spots in the zero-order Laue zone, which was indicative of Si-C bonds remaining between the SiC substrate and the buffer layer even after the Ca intercalation process. Thus, the interlayer distance between the buffer layer and the SiC substrate was determined to be fixed at the value of pristine graphene/SiC: 2.16 Å [17]. After the measurements, the sample was heated up to and kept at ~900°C to desorb the intercalated Ca atoms, whereupon the Ca-associated √3×√3 spots disappeared, as shown in Fig. 1 (c) (the spot intensity on the whole pattern became weaker than that of the pristine sample).

The transport property was measured of the Ca-intercalated bilayer graphene on SiC, prepared in the same way as mentioned above in a separate UHV system, using the *in-situ* four-point probe technique (Unisoku USM-1300S) [21]. As shown in Fig. 1 (e), the resistance dropped steeply at 4 K and reached zero at 2.2 K. The result clearly showed that the Ca-intercalated bilayer graphene has superconducting transition, as reported by Ichinokura *et al.* [5].

TRHEPD experiments were performed at the Slow Positron Facility (SPF) in IMSS, KEK, where a linac-based brightness-enhanced high-intensity positron beam was employed [16]. TRHEPD is a highly surface-sensitive technique because it utilizes the property of positive electrostatic potential in every material [22, 23]. The positron beam undergoes total-reflection at glancing angles smaller than a certain critical value, which in the case of graphene is approximately 2° for a 10-keV positron beam [24]. The positrons do not penetrate into the bulk under this condition; the diffraction intensity therefore depends entirely on the atomic arrangement of the topmost atomic layer on the surface. The rocking-curve, which here is (0 0) spot diffraction intensity measured as a function of the glancing angle ($\theta$), was extracted from a series of the TRHEPD patterns. The glancing angle was varied from 0.5° to ~6° with a 0.1° step by rotating the sample. The sample size was 15 mm×2 mm, and the positron beam size was less than Φ2 mm.

As shown in Fig. 1(f), we conducted the structural analysis of the data obtained with the positron beam in two azimuthal directions of incidence, the [1$\bar{1}$00] and 7.5° off the [1$\bar{1}$00]. The rocking-curve of the (0 0) spot in the "one-beam" condition (7.5° off the [1$\bar{1}$00] incidence) essentially includes the information of spacing in the out-of-plane (<0001>) direction only, while the rocking curve of the (0 0) spot in the "many-beam" condition (the [1$\bar{1}$00] incidence) includes information on the in-plane structure (on the (0001) plane) as well as on the out-of-plane structure. This is because the contribution of in-plane diffractions is significantly suppressed under the one-beam condition [25].

We first analyzed the rocking-curve in the one-beam condition to obtain the out-of-plane structure (interlayer distances), and then examined the rocking-curve in the many-beam condition to determine the in-plane structure with the obtained out-of-plane coordinates fixed. We minimized the difference between the measured rocking curves and the calculated rocking curves by optimizing the assumed structures, using a reliability factor $R$ [22, 26] as an index for degree of fit. The R-factor is defined as

$$R = \sqrt{\sum_i [I_{\text{exp}}(\theta_i) - I_{\text{cal}}(\theta_i)]^2} \times 100 \ (\%) \tag{1}$$

where $I_{\text{exp}}(\theta_i)$ and $I_{\text{cal}}(\theta_i)$ are the intensities, respectively, of the experimental and calculated diffraction spots at glancing angle $\theta_i$, with a normalization of

$$\sum_i I_{\text{exp}}(\theta_i) = \sum_i I_{\text{cal}}(\theta_i) = 1. \tag{2}$$

It has been established that Ca-intercalation into bulk graphite transforms the stacking of graphite layers from AB to AA type [18]; every second graphite layer is shifted to the [11$\bar{2}$0] direction by 1/√3 of the lattice constant in AB stacking, while all graphite layers are stacked without shift in AA stacking, as shown Fig. 1(f). As an analogue, a similar change in the stacking structure is expected for the bilayer graphene with Ca-intercalation. The "many-beam" condition with the incident beam directed along the [1$\bar{1}$00] is sensitive to the difference in the stacking of graphene layers because the in-plane structure perpendicular to the beam incidence direction is different between the two types of stacking.

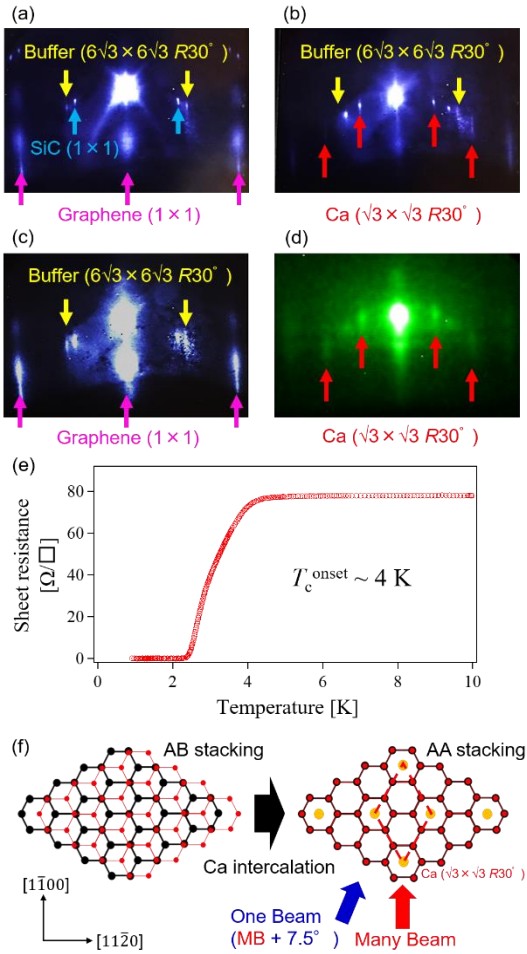

Fig. 1 RHEED patterns of (a) pristine bilayer graphene/SiC(0001), (b) Ca-intercalated graphene and (c) Ca-desorbed graphene. (d) TRHEPD pattern from Ca-intercalated graphene. (e) Temperature dependence of sheet resistance in Ca-intercalated graphene. (f) Simple schematic view of bilayer graphene and Ca-intercalated bilayer graphene without the buffer layer and SiC substrate. Shown here is the expected Ca-intercalation

change in the stacking structure from AB to AA type, in analogy with that from AB (graphite) to AA (bulk $C_6Ca$) stacking. Beam directions of the one- and many-beam condition are depicted by the blue and red arrows, respectively.

## 3. Results and discussion

We determined the structures of pristine bilayer graphene grown on SiC(0001), Ca-intercalated bilayer graphene and Ca-desorbed bilayer graphene. The bilayer graphene on SiC (0001) consisted of three carbon atomic layers on the substrate: two graphene sheets with a buffer layer below, as shown schematically in Fig. 2 (a). For the Ca-intercalated bilayer graphene on SiC, we checked three possible structural models, shown schematically in Figs. 2(b)-(d). They included those where Ca atoms were intercalated in the graphene-graphene interlayer [Fig. 2(b): model 1], the graphene-buffer interlayer [Fig. 2(c): model 2], and both interlayers [Fig. 2(d): model 3]. Model 1 had been thought to be the structure of Ca-intercalated bilayer graphene by analogy with the structure of bulk GIC, $CaC_6$.

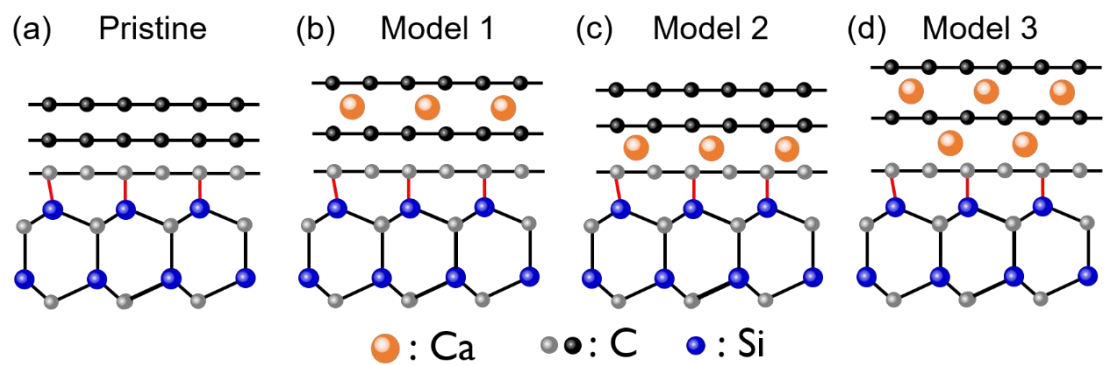

Fig. 2 (a-d) Schematic views of the structural models for the calculated curves. No attention has been paid to indicate the stacking of the buffer layer and the bilayer graphene.

First, the rocking-curves under the one-beam condition were analyzed. Shown in Fig. 3 are the raw experimental rocking-curves and the calculated rocking curves which were corrected for the effective sample size by multiplying with sin $\theta$ ($\theta$ is the glancing angle). This analysis yielded the interlayer distances with their associated error values [27]. The buffer layer was taken to be the same as a graphene sheet since contribution from this depth in the TRHEPD rocking curve is much lower than from a surface layer [22] and, additionally, the atomic arrangement is almost the same as graphene (its main structural difference with graphene is the presence of buckling, the magnitude of which is less than ~0.4 Å) [17, 19, 28]. In the case of pristine graphene, we analyzed the TRHEPD rocking curve with varying occupation parameters for 3 layers of graphene to explore the probable numbers of the layers present. The occupation of the carbon atoms in the graphene layers was included in the fitting parameters, ranging 0.7-1.0 ML for the bottom graphene layer, 0.7-1.0 ML for the middle graphene layer, and 0.0-0.3 ML for the top graphene layer. The ranges were set as such because the most likely layer number was already known to be bilayer from the ARPES measurement. The fitting yielded the occupation parameters of the bottom, middle, and top graphene layers as 1.00 ML, 0.97 ML, and 0.03 ML, respectively, supporting the findings of the ARPES measurements. The value of the R-factor was 0.76 %, which meant that most of the area (~94 %) was bilayer. The interlayer distances are obtained as: the bottom graphene and buffer layer 3.18 ± 0.14 Å, the middle and bottom graphene 3.38 ± 0.06 Å, and the top and middle graphene 3.66 ± 0.63 Å. These values are consistent with those previously derived from photoelectron diffraction [28] and X-ray diffraction [29] – graphene-buffer 3.24 ± 0.20 Å and 3.40 ± 0.10 Å, graphene-graphene 3.48 ± 0.10 Å and 3.40 ± 0.05 Å, respectively, within a typical analysis error of < ~0.2 Å [24, 27]. Hereafter in the analysis, the graphene component is considered to be a bilayer with some roughness induced by the Ca-intercalation and desorption process in the topmost layer obscuring possible contributions of small mono- and trilayer graphene areas in the rocking curve.

Fig. 3 (c) shows the rocking curves of the Ca-intercalated bilayer graphene under the one-beam condition. The blue (dashed), black (solid) and green (chain) lines are based on models 1-3 [Figs. 2(b)-(d)] with structure optimization, respectively. All of the calculated curves commonly deviate from the experimental data in the total reflection region (glancing angle < 2°), which suggests some roughness of the topmost surface caused by the Ca-intercalation. Therefore, we calculated the R-factor with data outside the total reflection region to evaluate the structures of the Ca-intercalated and the desorbed bilayer graphene. The black solid line (model 2) in Fig. 3(c) has the best agreement with the experimental curve ($R$ = 1.43 % for the higher glancing angle > 2°), which indicates that the Ca atoms are intercalated in the graphene-buffer interlayer only. The interlayer distances obtained are shown in Fig. 3(d); the graphene-buffer spacing is 4.21 ± 0.11 Å and that of graphene-graphene is 3.33 ± 0.16 Å. The Ca atoms are located in the graphene-buffer interlayer, 1.46 ± 2.24 Å above the buffer layer. The Ca-intercalated interlayer-distance is close to that of the bulk $CaC_6$ (4.5 Å [18]). The non-Ca-intercalated interlayer-distance remains almost the same as in the pristine graphene.

The open circles in Fig. 3(e) show the measured rocking-curve under the one-beam condition for the Ca-desorbed bilayer graphene. Its shape is similar to, but slightly different from the measured rocking curve for the pristine bilayer graphene (Fig. 3 (a)). Lower values in the total reflection region indicate some disorder in the top graphene layer. The slight change in the shape in the higher glancing-angle region > 2° indicates some change in the interlayer distances. In fact, our analysis showed a larger graphene-buffer interlayer distance than the pristine one. The obtained interlayer distances are noted in Fig. 3(f): the graphene-buffer distance is 3.84 ± 0.14 Å and the graphene-graphene distance is 3.35 ± 0.21 Å ($R$ = 1.15 % for the higher glancing-angle region > 2°). This result indicates that the Ca-desorption after the intercalation induces an irreversible structure change - an expansion of the interlayer distance.

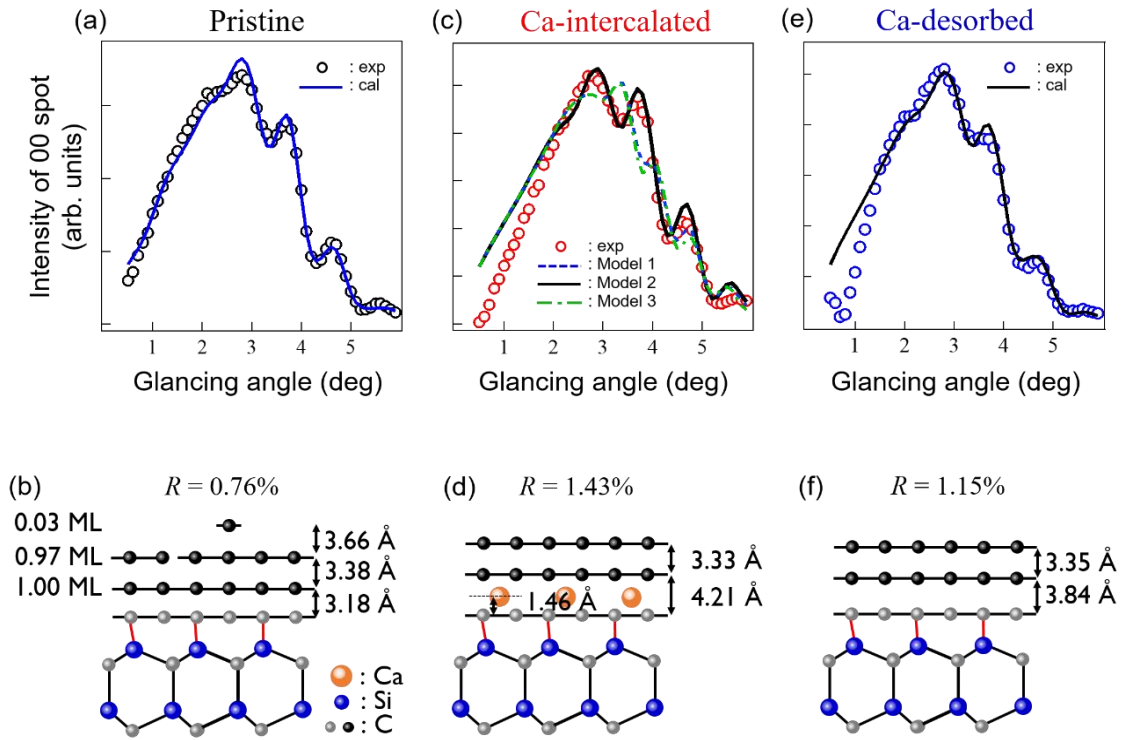

Fig. 3 TRHEPD rocking curves under the one-beam condition with calculated curves for (a) pristine bilayer graphene, (c) Ca-intercalated graphene and (e) Ca-desorbed graphene in the one-beam condition. Schematic views of the structural analysis results for (b) pristine bilayer graphene/SiC(0001), (d) Ca-intercalated graphene and (f) Ca-desorbed

graphene. No indication has been made of the stacking of the buffer layer and the bilayer graphene since the results in the one-beam condition are not sensitive to it.

In the analysis, the mean-thermal-vibration amplitudes of graphene, buffer and Ca layers were included in the fitting parameters. The optimized values are: $0.145 \pm 0.013$ Å for graphene and the buffer layers in pristine bilayer graphene, $0.065 \pm 0.046$ Å for graphene and the buffer layers and $0.276 \pm 0.276$ Å for the Ca layer in Ca-intercalated graphene, and $0.142 \pm 0.039$ Å for graphene and the buffer layers in Ca-desorbed graphene.

Next, rocking-curves under the many-beam condition were analyzed. Fig. 4(a) shows the measured and calculated rocking-curves for the pristine bilayer graphene. The calculated curves were optimized for AAA(A) stacking ($R = 3.84$ %) and for ABA(B) stacking ($R = 0.93$ %), both including the buffer layer (The stacking is indicated from buffer layer to top graphene layer). The distinction is clear at angles larger than 4° - the ABA(B) stacking model is a much better fit than the AAA(A) stacking model to the experimental data, which is consistent with an energetically stable structure [30]. The assumed AAA(A) stacking structure in the bilayer graphene produces a much higher intensity than the ABA(B) stacking at the glancing angle of 4-5° in the calculation. This is because the positron waves scattered from the two graphene layers interfere with each other constructively in the case of AAA(A) stacking whilst destructively in the ABA(B) stacking due to the lateral shift of the graphene lattice. In this way, we can clearly distinguish the stacking structure by the many-beam-condition experiments.

The experimental rocking curve for the Ca-intercalated bilayer graphene under the many-beam condition was compared to the calculated curves based on different stacking structures [Fig. 4(b)]. We denote the stacking sequence including the location of Ca atoms from the bottom to top layers; for example, "A-Ca-BA" indicates that the buffer layer is A, the middle graphene is B, the top graphene is A, and Ca atoms are intercalated in the graphene-buffer interlayer. The Ca atoms locate at the center of hexagonal lattices of the buffer layer because the distance from the buffer layer to Ca atom is closer than from the graphene layer above (otherwise the overlap between Ca and C atoms makes the structure energetically unstable). The "A-Ca-BA" ($R = 1.70$ %) stacking sequence was found to give the best agreement with the experimental data, being more plausible than the "A-Ca-AB" ($R = 2.07$ %) stacking. The result suggested that no stacking shift from the pristine one occurred due to the Ca-intercalation unlike in the case of bulk graphite [18]. The length between a Ca atom and the carbon atom of the graphene layer (which is just above the Ca atom) is 2.75 Å and that between a Ca atom the carbon atoms of the buffer layer is 2.04 Å. While the former length is close to the Ca-C length of bulk $CaC_6$ (2.66 Å), the latter is considerably smaller. This deviation may be related to differences in the structure (stacking structures, interlayer distances and Si-C bond with buffer layer etc.).

Fig. 4(c) shows the measured and optimized rocking-curves in the many beam condition for the Ca-desorbed bilayer graphene. The calculated curves were based on the optimized model of ABA stacking bilayer graphene including the buffer layer shown in Fig. 2(a) as for the pristine bilayer graphene. The optimized rocking curve agrees well with the experimental data at a glancing angle larger than 2°.

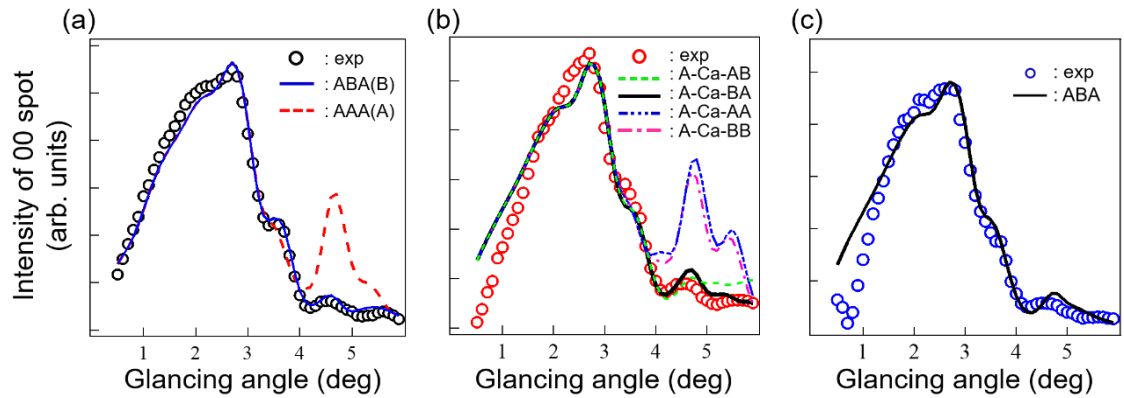

Fig. 4 TRHEPD rocking curves under the many-beam condition with the calculated curves for (a) pristine bilayer graphene, (b) Ca-intercalated graphene and (c) Ca-desorbed graphene in the many-beam condition.

We have summarized the structures determined in this study of pristine bilayer graphene on SiC(0001), Ca-intercalated bilayer graphene and Ca-desorbed bilayer graphene in Fig. 5(a)-(c), respectively. The results show clearly that the Ca-atoms are intercalated between the graphene layer and the buffer layer, and not between the two graphene layers. This is a significant step in elucidating the superconducting mechanism of Ca-intercalated bilayer graphene on SiC. To further detail the superconducting mechanism of this compound, the band calculation for our structural model is needed. Furthermore, this investigation suggests the possibility of superconductivity in monolayer graphene on SiC since it has been found that the contribution of the top graphene layer to this property may be negligible. It is also expected that Ca-intercalated monolayer graphene can be a system where superconductivity and Dirac electrons can coexist.

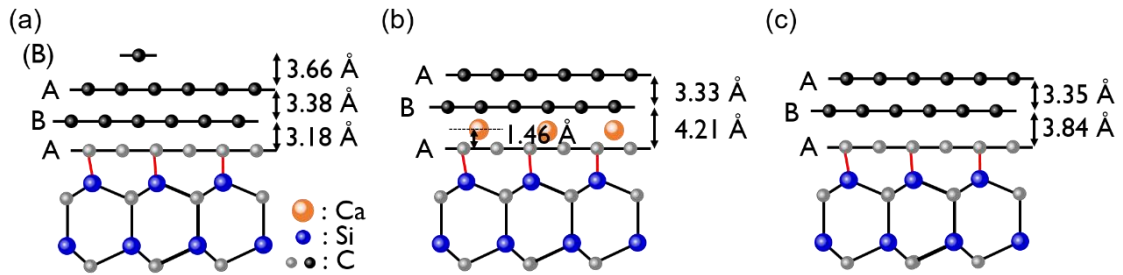

Fig. 5 Schematic views of the structure determined for (a) pristine bilayer graphene/SiC(0001), (b) Ca-intercalated graphene and (c) Ca-desorbed graphene. The stacking determined for the buffer layer and the bilayer graphene are shown schematically and also by the symbols A and B.

## 4. Conclusion

In conclusion, a structural analysis of superconducting Ca-intercalated bilayer graphene on SiC(0001) substrate was performed using TRHEPD. The results clearly show that the Ca atoms are intercalated only between the graphene layer and the buffer layer. This is the first elucidation of the atomic arrangement of superconducting Ca-intercalated bilayer graphene.

**ACKNOWLEDGEMENTS**

We thank T. Iimori (Institute for Solid State Physics, The University of Tokyo) and M. Imamura (Synchrotron Light Application Center, Saga University) for their useful discussion and assistance in experiments to determine the number of graphene layers by ARPES study. This study was performed under the PF Proposal No. 2017G519 at Institute of Materials Structure Science, High Energy Accelerator Research Organization (KEK).

# Supplementary Data

# Structure of Superconducting Ca-intercalated Bilayer Graphene/SiC studied using Total-Reflection High-Energy Positron Diffraction

Y. Endo*[1], Y. Fukaya[2], I. Mochizuki[3], A. Takayama[4], T. Hyodo[3] and S. Hasegawa[1]

[1] Department of Physics, School of Science, The University of Tokyo,

7-3-1 Hongo, Bunkyo-ku, Tokyo 113-0033, Japan

[2] Advanced Science Research Center, Japan Atomic Energy Agency,

2-4 Shirakata, Tokai, Naka, Ibaraki 319-1195, Japan

[3] Institute of Materials Structure Science, High Energy Accelerator Research Organization (KEK), Oho 1-1, Tsukuba, Ibaraki 305-0801, Japan

[4] Department of Physics, School of Advanced Science and Engineering, Waseda University,

3-4-1 Okubo, Shinjuku-ku, Tokyo 169-8555, Japan

*Corresponding author. Tel: +81-3-5841-4209. E-mail: y.endo@surface.phys.s.u-tokyo.ac.jp
(Yukihiro Endo)

## 1. Band dispersion of the pristine graphene

We performed ARPES measurements to obtain the band dispersion of the pristine graphene sample. Figure S1 shows the band dispersions around the K point in the Brillouin zone. We found two parabolic π bands, which is characteristic of bilayer graphene. As expected for epitaxial bilayer graphene on SiC (0001), the Fermi level is shifted by around 0.3 eV above the Dirac energy of the π bands [5, 15]. This result indicates that the dominant area of the sample is bilayer.

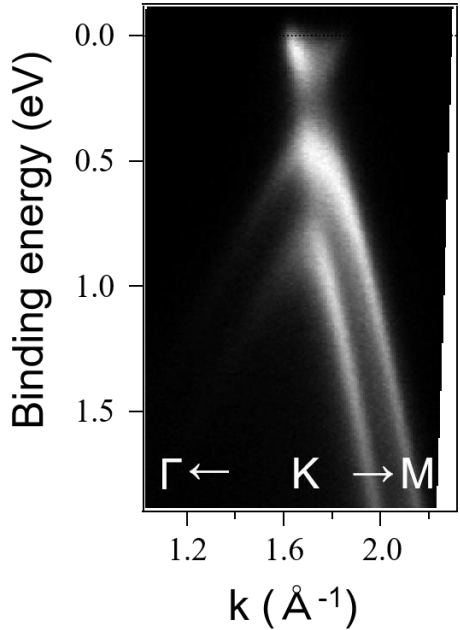

**Figure S1.** Observed band dispersion along the Γ-K-M direction around the K point of pristine graphene.